\documentclass[conference]{IEEEtran}
\IEEEoverridecommandlockouts
\usepackage{cite}
\usepackage{amsmath,amssymb,amsfonts}
\usepackage{algorithmic}
\usepackage{algorithm}
\usepackage{graphicx}
\usepackage{subfigure}
\usepackage{textcomp}
\usepackage{xcolor}
\def\BibTeX{{\rm B\kern-.05em{\sc i\kern-.025em b}\kern-.08em
    T\kern-.1667em\lower.7ex\hbox{E}\kern-.125emX}}

\begin{document}

\title{DistrEE: Distributed Early Exit of Deep Neural Network Inference on Edge Devices\\
}

\author{\IEEEauthorblockN{Xian Peng\IEEEauthorrefmark{1},
Xin Wu\IEEEauthorrefmark{1}, Lianming Xu\IEEEauthorrefmark{2}, Li Wang\IEEEauthorrefmark{1} and Aiguo Fei\IEEEauthorrefmark{1} 
}
\centerline{\IEEEauthorrefmark{1}
School of Computer Science (National Pilot Software Engineering School), }\\ 
{Beijing University of Posts and Telecommunications, Beijing, China}\\
 \centerline{\IEEEauthorrefmark{2}
 School of Electronic Engineering, Beijing University of Posts and Telecommunications, Beijing, China}\\
\\





\thanks{This work was supported in part by the National Natural Science Foundation of China under Grant U2066201, 62171054, and 62101045, and in part by the Natural Science Foundation of Beijing Municipality under Grant L222041, in part by the Fundamental Research Funds for the Central Universities under Grant No. 24820232023YQTD01, 2024RC06, and 2023RC96, in part by the Double First-Class Interdisciplinary Team Project Funds 2023SYLTD06. (Corresponding author: \emph{Li Wang}.)}
}

\maketitle

\begin{abstract}
Distributed DNN inference is becoming increasingly important as the demand for intelligent services at the network edge grows. By leveraging the power of distributed computing, edge devices can perform complicated and resource-hungry inference tasks previously only possible on powerful servers, enabling new applications in areas such as autonomous vehicles, industrial automation, and smart homes. However, it is challenging to achieve accurate and efficient distributed edge inference due to the fluctuating nature of the actual resources of the devices and the processing difficulty of the input data. In this work, we propose DistrEE, a distributed DNN inference framework that can exit model inference early to meet specific quality of service requirements. In particular, the framework firstly integrates model early exit and distributed inference for multi-node collaborative inferencing scenarios. Furthermore, it designs an early exit policy to control when the model inference terminates. Extensive simulation results demonstrate that DistrEE can efficiently realize efficient collaborative inference, achieving an effective trade-off between inference latency and accuracy.

\end{abstract}


\section{Introduction}

Deep Neural Networks (DNNs \cite{hong2024spectralgpt}) that are required for high accuracy often have deeper or wider architectures, leading to increased memory occupation and computational costs. This impedes the deployment of DNNs on resource-constrained mobile or Internet of Things (IoT) devices. Currently, there are many methods used to reduce the computational overhead of DNNs to provide efficient edge intelligence services, including model compression and inference offloading \cite{ai-emergency}. However, these methods can excessively compress the network, resulting in a decrease in inference accuracy, and they heavily rely on unreliable wide area network connections \cite{review}. The distributed DNN inference paradigm can allocate the workload of DNNs across multiple devices to help alleviate the computational burden on a single edge device and fully utilize the available resources on nearby devices. Moreover, keeping data local and performing inference nearby not only protects privacy but also avoids unpredictable delays caused by remote transmission. However, during the inference process, the actual available resources of the device and the processing difficulty of the input data fluctuate, and not all input data require the same amount of computation to produce inference results that meet the quality of service requirements. Performing the same model computation on all input data is highly detrimental to improving inference performance \cite{adaptiveNN}, especially in edge computing scenarios with limited device resources. Applying the same model to both simple and complex data leads to resource wastage and increased latency. 

Based on the fact that not all inputs require the same amount of computation to produce a reliable prediction, previous works explored a wide range of techniques to achieve dynamic inference. Dynamic inference techniques aim to adapt to different input data, service requirements, and resource fluctuations by dynamically adjusting the inference plan. Teerapittayanon et al. in \cite{branchynet} propose BranchyNet, which realizes efficient inference by introducing early exit branches at middle layers of the DNN and selectively exiting the network early depending on the input instance. AdaEE \cite{adaEE} is proposed to avoid the limitation of manually set exit rules in BranchyNet, which dynamically adjusts inference confidence thresholds based on context. AdaEE estimates the accuracy and computational overhead of each branch based on the historical inference results and the target accuracy requirements and adjusts the early exit decision accordingly, which can strike a balance between computational latency and accuracy. Another class of methods that dynamically omit local model computations includes skipping certain convolutional layers or network blocks \cite{blockdrop} to achieve adaptive inference computations. These methods often utilize trainable gating components within the network architecture, leading to a complex training process and limiting flexibility in deployment across different hardware platforms. 
Dynamic selection of model variants is a simpler approach to dynamic inference, EdgeAdaptor \cite{edgeadpatpr} trains a series of models with different latency and accuracy specifications and deploys all of them on the target device. At inference time, the most appropriate model for each input is selected to perform inference based on a specific rule.

All of the above methods are dynamic inference techniques proposed for single-node model inference, and some works investigated dynamic inference methods for multi-node collaborative inference scenarios.
For device-edge collaborative inference, Edgent \cite{edgent} combines model splitting with on-demand early exit to utilize the computational resources of end devices and edge servers, and dynamically adjusts the splitting and execution of inference according to the actual latency requirements to achieve more efficient and flexible inference. 
For collaborative inference based on input partitioning, CoEdge \cite{coedge} adjusts the division of inputs at each layer according to real-time computing resources and network conditions to achieve adaptive and efficient inference and improve efficiency and resource utilization. However, this method can only adapt to resource fluctuations, and cannot adapt to input difficulty variability to save computational overhead. In the collaborative inference scenario based on the knowledge distillation division model (NoNN) \cite{nonn}, multiple nodes deploy functionally independent sub-models, and the existing dynamic inference techniques cannot be directly applied, thereby dynamic collaborative inference methods for this scenario still need to be studied. The NoNN introduces memory and communication-aware model compression for distributed model inference on IoT devices. Still, it fails to differentiate between predicting complex and simple images and thus fails to adjust inference plans to improve performance.

To this end, we propose an efficient adaptive inference network based on an early-exit mechanism, called DistrEE. Our key contributions are summarized as follows: 

\begin{itemize}
\item We present an efficient distributed DNN inference framework, called DistrEE. It combines distributed inference with early exit to achieve adaptive inference in dynamic scenarios. 
\item We design an effective end-to-end loss function for the joint training of distributed neural network models and early exits. 
\item We evaluate the performance of DistrEE via extensive simulations, verifying the effectiveness of the proposed training framework and the superior performance of adaptive inference in terms of latency-accuracy trade-off.
\end{itemize}

\section{Background Knowledge}

\subsection{Network of Neural Networks (NoNN)}
To achieve efficient edge inference, distributed DNN inference became an important and promising solution. Zeng et al. \cite{coedge} proposed a co-inference approach CoEdge, where the input of each layer is used as a unit for workload distribution. In this case, an intermediate feature map that is located in the boundary of partitioning should be transferred, resulting in considerable communication burdens. To get rid of such prohibitive communication overheads, Bhardwaj et al. \cite{nonn} proposed a completely different distribution approach, called Network
of Neural Network (NoNN), based on Knowledge Distillation (KD). In NoNN,  the original model was partitioned into multiple independent and smaller models. Specifically, NoNN focused on the final convolutional (fconv) layer of a given trained model, referred to as the teacher network, the filters in fconv were grouped into multiple exclusive subsets to derive student models. Each filter in fconv
contributed differently to the inference result, and dividing the filters should take into account the correlation between individual filters. 
To determine how the filters should be grouped, NoNN proposed a network science-based solution. NoNN first constructed an undirected graph where each vertex represents a filter in fconv, and the correlation between a pair of filters was quantified and assigned as the weight of corresponding edge in the graph. Formally, the weight of the edge between $i$ and $j$ is determined as follows: 

\begin{equation}\label{AH}
\begin{aligned}
    w(i,j)= \sum_{val} a_i a_j|a_m-a_j|,
\end{aligned}
\end{equation}
where $a_i$ is the average activation of filter $i$. In the NoNN approach, filters that respond simultaneously were encouraged to be evenly distributed across multiple student networks because they want to distribute knowledge as evenly as possible to all students. 
After this, a student network was trained to learn certain knowledge from each subset of the partitioned filters. The training loss of NoNN is defined as follows:

\begin{equation}\label{lossNonn}
\begin{aligned}
    Loss(\theta_S)= Loss^{kd}(\theta_S) + \beta \sum_{P \in fconv} Loss^{act}(\theta_S,P).
\end{aligned}
\end{equation}

In the inference stage, each student network was executed independently on a single device and the results were collected only once at the end to infer the final prediction. In this way, the heavy communication overheads caused by other distributed inference approaches \cite{coedge} can be mitigated.

\subsection{Early Exit}
Based on the observation that features learned in the early stages of a deep network tend to infer the majority of the data correctly \cite{hong2024multimodal}, early exits were proposed to allow some samples to execute just a portion of the DNN. Kung et al. \cite{branchynet} proposed the concept of BranchyNet for exiting early in inference to save time and resources. The BranchyNets are DNNs with side branches inserted in the intermediate layers so that simpler input samples can be early inferred and, hence, reduce the inference time. 
In BranchyNet, each side branch consists of several convolutional layers and a fully connected (FC) layer, where the FC layer produces a vector $z$. Afterwards, a sotmax operation is performed on $z$ to generate a probability vector $\hat{y}$, where each element represents the probability of the sample in each class. 

In the inference phase, the confidence level is estimated based on the value of $\hat{y}$ generated at each exit to decide whether the inference exits early or not. BranchyNet used entropy as the measure to quantify the 
prediction confidence at each exit point, the entropy is defined as:

\begin{equation}\label{entropy}
\begin{aligned}
    entropy(\hat{y})= -\sum_{y_c \in \hat{y}} y_c \log y_c.
\end{aligned}
\end{equation}

High entropy indicates similar probabilities across classes and therefore a non-confident prediction, while low entropy suggests a single peak result. The entropy value was compared with a preset threshold and if the entropy value is below, the inference is stopped and the output is returned.

\section{DistrEE Design}
In this section, we present the detailed design of the DistrEE scheme, which aims to combine distributed inference with early exit to achieve adaptive inference in dynamic scenarios. Our scheme is divided into two stages. The first stage focuses on offline joint training of neural networks with multiple branches, while the second stage investigates early exit strategies for online dynamic inference.

\begin{figure}[t]
    \centering
    \includegraphics[width=0.48\textwidth]{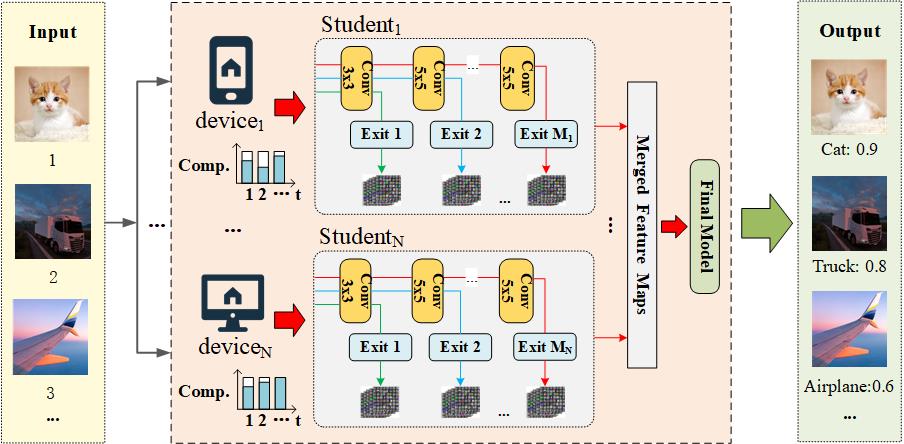}
    \caption{Inference scenario of DistrEE.}
    \label{sce}
\end{figure}

\subsection{Architecture}
To develop an inference plan that meets the task requirements for diverse input data and fluctuating device resources to achieve efficient dynamic collaborative inference, we propose DistrEE, a dynamic collaborative inference framework based on multi-branch neural networks. According to the execution phase, the dynamic inference method based on multi-branch neural networks can be divided into two steps: joint training of multi-branch models and online decision-making for dynamic inference. 
Consider a scenario where an edge cluster consisting of multiple edge devices collaborates to perform DNN inference, the system scenario is shown in Fig. \ref{sce}. Given a set of edge devices $D=\{d_1,d_2,...,d_N\}$ and an original teacher model $T$, the student models obtained by distillation of the teacher model are denoted as a unified student model $\mathcal{S}=\{S_1, S_2,..., S_N\}$. In the student model $S_i$, there are $M_i$ different exit branches in the backbone DNN network, which divides the DNN into $M_i$ sequentially connected stages. The network structure of the exit branches after each stage is determined based on the structure of the backbone DNN output layer to generate intermediate feature maps of the same shape.
For a given input data, when the intermediate features obtained in a branch are sufficient for a reliable prediction, it can exit from that branch early to save the device's computational resources and accelerate the inference. Input that is not exited early ends the inference at the last backbone exit of the model.

\begin{figure}[t]
    \centering
    \includegraphics[width=0.48\textwidth]{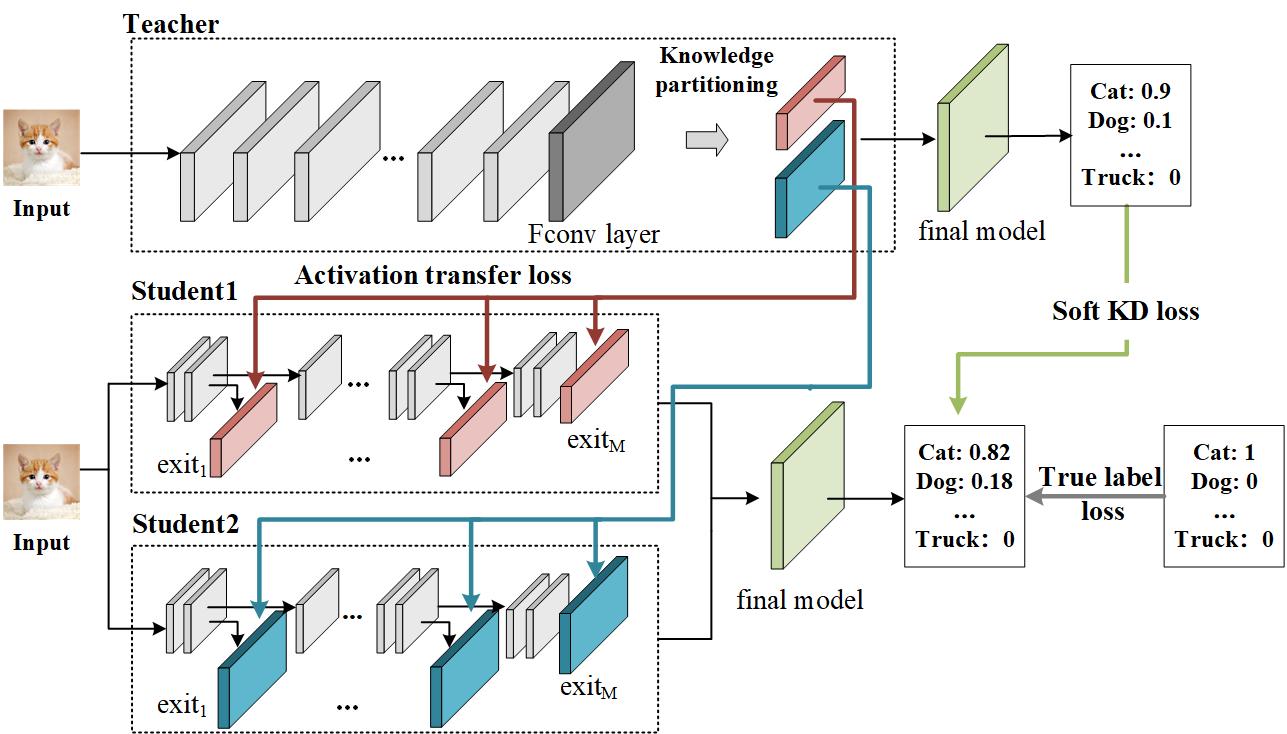}
    \caption{Joint training with multiple branches.}
    \label{train}
\end{figure}

\subsection{Joint Training}
DistrEE assumes that all student models have the same backbone DNN, which is reasonable since networks with multiple branches are inherently adapted to scenarios where devices are heterogeneous. Specifically, for a weaker device, it is sufficient to split only a portion of the DNN in front of a particular exit for deployment.
Therefore, we establish $M_i=M, \forall i\in\{1,...,N\} $, as depicted in Fig.\ref{train}. 

The objective of joint training multi-branch student models is to enable each branch of each student model to effectively learn from the teacher model, allowing as many inputs as possible to exit inference at an early exit. For the branch $j$ of the unified student model, when it undergoes training as an independent entity, the same training loss function utilized in NoNN is applicable, as delineated in Eq. \ref{KDloss}. In our framework, the improved end-to-end training method for the multi-branch neural network model is used to train the unified student model. The end-to-end training loss function is given by Eq. \ref{lossSum}, which represents the weighted sum of the training losses corresponding to each branch.

\begin{equation}\label{KDloss}
\begin{aligned}
    Loss(\theta_S,j)= & \begin{matrix}\underbrace{(1-\alpha)\mathbb{H}(y,P_{S,j})+\alpha\mathbb{H}(P_T^\tau,P_{S,j}^\tau)}\\\text{KD loss}\end{matrix} \\ 
    & +\begin{matrix}\underbrace{\beta \sum \limits_{P \in fconv} \left \| \frac{v_T^F(P)}{||v_T^F(P)||}-\frac{v_{S,j}^F(P)}{||v_{S,j}^F(P)||} \right\|^2}\\\text{AT loss}\end{matrix},
\end{aligned}
\end{equation}

\begin{equation}\label{lossSum}
\begin{aligned}
    Loss(\theta_S)= \sum_{j=1}^M w_j Loss(\theta_S,j),
\end{aligned}
\end{equation}
where $\theta_S$ denotes the parameters of the student network, $\mathbb{H}$ is the standard cross-entropy loss. The first term represents the standard knowledge distillation loss that combines the hard and soft-label cross-entropy losses. The second term corresponds to the AT loss, which measures the error between the output feature maps of the fconv partition of the teacher and the output feature maps of the branch $j$ of the corresponding student. $w_j$ is used to control the loss weight of the exit branch $j$ of the unified student model. Based on the above loss function, the classical stochastic gradient descent method can be used for end-to-end joint training of student models with multiple branches.

\subsection{Dynamic Inference}
Once training is complete, the student networks can be deployed to perform fast inference by predicting samples in the early stages of the network according to a specific exit strategy. If the feature maps observed at an exit branch contains sufficient information about the input sample, the inference exits and returns an intermediate feature early, and computation in subsequent branches of the network can be omitted. Therefore, how to measure the confidence level of intermediate features is crucial to achieve efficient early exit. BranchyNet \cite{branchynet} uses a threshold on the entropy of the softmax predictions as a measure of confidence. For each exit point, the input sample is fed through the corresponding branch. If the entropy is less than the given threshold, the class label with the maximum score is returned. Other approaches directly use the top-1 softmax value as a quantification of confidence and inference exits when the top-1 value is large enough. 

\begin{figure}[!t]
\renewcommand{\algorithmicrequire}
{ \textbf{Input:}}
\renewcommand{\algorithmicensure}
{ \textbf{Output:}}
\begin{algorithm}[H]
\caption{Dynamic Inference Algorithm}
\label{alg1}
\begin{algorithmic}[1]
\REQUIRE Difference threshold vector $T=[t_1,t_2,...,t_M]$.
\ENSURE Feature map of selected exit.
\STATE Feed input data into the student network
\FOR{$j=1 \rightarrow M$}
    \STATE Compute $diff_{cos}(F_j,F_1)$
    \IF{$diff_{cos}(F_j,F_1) > t_j$}
    \RETURN $F_j$
    \ELSE 
    \STATE Continue to perform forward propagation.
    \ENDIF
\ENDFOR
\RETURN $F_M$
\end{algorithmic}
\end{algorithm}
\end{figure}

\begin{figure}[t]
\centering
\subfigure[The similarities between neighboring exits]{ 
    \includegraphics[width=0.45\textwidth]{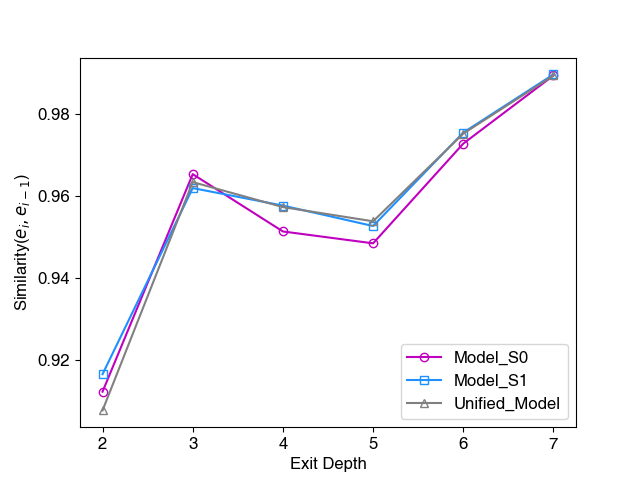}} 
\hspace{1in} 
\subfigure[The difference relative to first exit]{
    \includegraphics[width=0.45\textwidth]{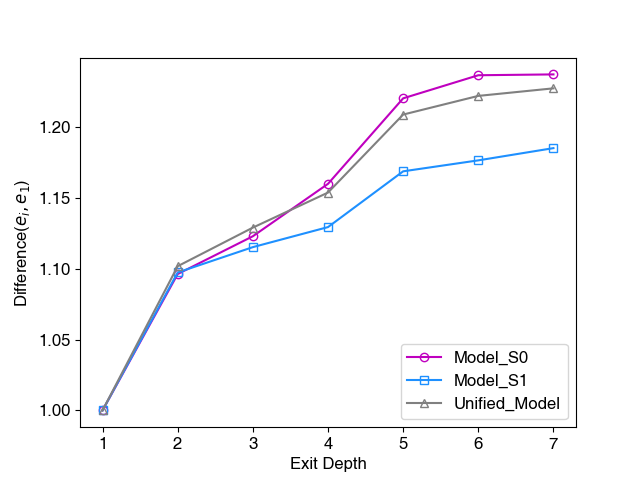}}
\caption{Two criteria used to measure the level of confidence.}
\label{rules}
\end{figure}

However, the outputs of individual student models in DistrEE are only concerned with local feature information, and the outputs of all student models need to be aggregated and then computed by the final inference model related to the task to obtain the complete inference results. Therefore, traditional methods based on confidence measures such as top-1 probability or entropy value are not applicable in this scenario, i.e., the output of a single student cannot be directly used to measure the softmax confidence. Recent works on the encoder-decoder framework face a similar problem because the hidden state representation of the encoder is not directly related to the final prediction. 
Tang et al. \cite{similarity} observe the hidden-state representation of each layer in Transformer will reach a saturation status in the language model, which indicates that the hidden-state representation change decreases as going through the latter layers. Therefore, the latter layers can be skipped safely without a significant performance drop when such saturation status is reached. Based on this observation, they leverage the Cosine Similarity between layers as a proxy for saturation level to control early exit. Inspired by that work, we investigate whether the saturation state exists in a multi-branch convolutional neural network. The similarity of the output features between two neighboring exits is shown in Fig.\ref{rules} (a), these results indicate that similar saturation states do not exist in our models.

On the contrary, we propose another criterion based on the degree of feature differentiation, which takes the output features of the shallowest exit as the benchmark and considers the first exit as incapable. In this way, the degree of difference in the output features relative to the output features of the first exit reflects the improvement in the learning ability of the corresponding exit, i.e., the greater the difference, the better the learning ability. Fig. \ref{rules} (b) shows the experimental results of exit feature differences, the degree of feature differences relative to the first exit increases with the depth of the exit. Based on this result, we use relative feature difference to measure the confidence level of each exit and propose a dynamic inference control algorithm based on feature difference, as shown in Algorithm \ref{alg1}. Where $F_j$ represents the output feature map of exit $j$, $diff_{cos}(F_j,F_1)$ is computed as Eq. \ref{diffcos}. When the observed feature difference at exit $j$ is greater than a given threshold $t_j$, inference terminates and the feature map $F_j$ is returned. The rationality of the criterion and the effectiveness of the algorithm are verified in Section \ref{evalution}.

\begin{equation} \label{diffcos}
    diff(F_j,F_1)=\frac{1}{sim_{cos}(F_j,F_1)}=\frac{\|F_j\|\cdot\|F_1\|}{F_j\cdot F_1}.
\end{equation}

\section{Performance Evaluation} \label{evalution}
In this section, we evaluate the performance of the proposed distributed inference scheme DistrEE via extensive simulations. Specifically, we verify the effectiveness of the training loss function used to jointly train multi-branch neural networks. Then, we compare the performance of the proposed dynamic inference algorithm with several baselines in terms of inference overhead and accuracy.

We use WideResNet-16×4, a well-trained classification network on the CIFAR-10 dataset, as the teacher model, which achieves a classification accuracy of 91.86\%. Unified student model DistrEE-2 contains 2 student models, and the student model uses WideResNet-16×1 as the backbone network. The WideResNet model is designed with a shortcut connection in the basic block structure, so appending early exit branches should be done in base block units. In DistrEE-2, we attach exit branches after the 1st, 4th, 6th, 9th, 11th, 14th, and 16th convolutional layers in the WideResNet-16×1 architecture. Notably, the exit following the 16th convolutional layer serves as the original backbone network's output exit. The architecture of each exit branch is designed as a cascade, including Relu and Avg-pool layers, consistent with the exit structure of the backbone network. The computational demands of both the backbone network and the exit structures for each branch in the DistrEE-2 setup are comprehensively detailed in Table \ref{comp}. The total computational overhead of a single student model in DistrEE-2 is 34.55 MFLOPs and the number of parameters totals 189044, while the total computational overhead of the backbone model WideResNet-16×1 is 34.26 MFLOPs and the number of parameters totals 178540. Compared to the original model WideResNet-16×1, the additional computational overhead and the number of model parameters introduced by the additional 6 early exit branches are only 0.88\% and 5.88\% of the original model, and thus do not result in a burdensome additional overhead.

\begin{table}[!htbp]
\renewcommand\arraystretch{1.5}
  \begin{center}
    \caption{Computational overheads of DistrEE-2.}
    \begin{tabular}{cccccccc}
    \hline
    Branch & E1 & E2 & E3 & E4 & E5 & E6 & E7
    \\
    \hline
    \begin{tabular}[c]{@{}c@{}} Backbone \\ (MFLOPs)\end{tabular} & 0.459& 4.817& 4.817& 7.356& 4.768& 7.28& 4.74\\
    \hline
    \begin{tabular}[c]{@{}c@{}} Exit \\ (MFLOPs)\end{tabular} & 0.067& 0.067& 0.067& 0.035& 0.035& 0.02& 0.02\\ \hline
    \end{tabular}
    \label{comp}
  \end{center}
\end{table}

\begin{figure}[t]
\centering
\subfigure[Training loss.]{ 
    \includegraphics[width=0.45\textwidth]{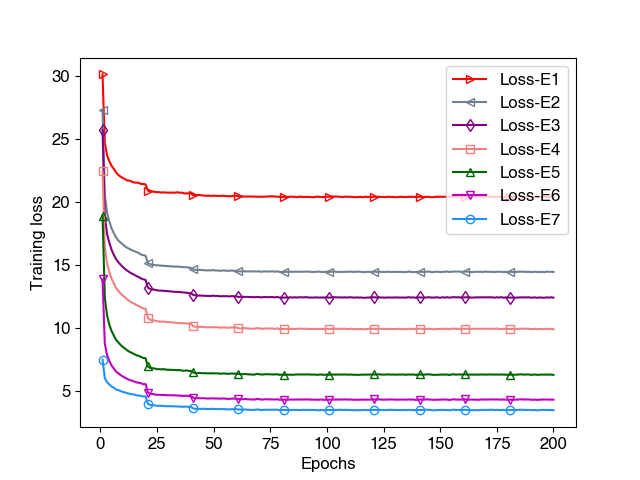}} 
\hspace{1in} 
\subfigure[Training accuracy.]{
    \includegraphics[width=0.45\textwidth]{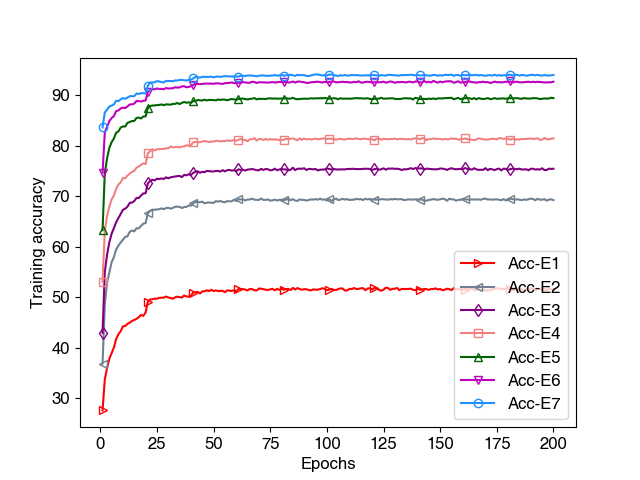}}
\caption{Training results for DistrEE-2.}
\label{resTrain}
\end{figure}

Using the training loss function expressed in Eq. \ref{lossSum}, we set the training loss weight vectors of the seven exit branches of DistrEE-2 to $w$ = [1, 1, 1.1, 1.4, 1.4, 1.3, 1.2], aiming at allowing the intermediate exits to learn efficiently and thus increase the chances of early exit. The model is trained based on the CIFAR-10 dataset with 200 epochs of training rounds, and the model training results are shown in Fig.\ref{resTrain}. The test accuracy of the seven branches of the student model on the test set is [47.16\%, 66.34\%, 72.96\%, 78.68\%, 86.96\%, 89.46\%, 90.57\%]. The results show that the proposed joint training framework for multi-branch models can train the student model effectively, and all branches of the model can reach the convergence state quickly.

To validate the performance of the proposed dynamic inference algorithm, images are randomly selected from the CIFAR-10 dataset to form a test dataset, which contains 10 images from each of the 10 categories, totaling 100 images. The trained model DistrEE-2 is used to execute inference based on different early exit policies, and the experiment is repeated 10 times on the 
Lenovo XiaoXin Pro 14
to record the average inference latency and accuracy for comparing the performance of different approaches. For DistrEE, we set the difference threshold vector as $T$ = [1, 1.12, 1.14, 1.16, 1.18, 1.20, 1.22]. Specifically, we compare the performance of our DistrEE approach to the following baseline:

\begin{enumerate}
    \item \textit{Last-exit}: Instead of a dynamic early exit, each input performs full model inference, exits at final exit E7, and returns inference results.
    \item \textit{Random}: Randomly selects the exit location from the 7 exits of the model for each input.
    \item \textit{Similarity-based}: For a given input, observe the output features at each exit, based on the cosine similarity of the output features of the two neighboring exits, and exit when the similarity is greater than a given threshold. The exit threshold vector used in the experiment is [0.97, 0.97, 0.97, 0.97, 0.97, 0.975, 0.98].
\end{enumerate}

The average performance of executing model collaborative inference using the above exit strategies is summarized as shown in Table \ref{performance}. The Last-exit approach uses the final exit directly without an early exit, thus its inference accuracy, computational overhead, and inference latency are all the highest. The Random exit method randomly selects exit location with almost the same selection probability for each exit, without considering observable information, and therefore has lower inference accuracy, computational overhead, and inference latency. The Similarity-based method decides to exit using the similarity of the output feature maps of the neighboring exits, and the experimental results show that the decision based on this similarity does not make a good decision, and leads to a significant loss of accuracy while reducing the computational overhead and the inference latency. DistrEE implements early exits using the Algorithm \ref{alg1} proposed in this paper, which is based on the difference in feature relative to the first exit. Algorithm \ref{alg1} assumes that the greater the degree of difference in the output feature of an exit relative to the first exit, the stronger the learning ability of that branch. In DistrEE, the degree of feature differences of the exits increases with the depth of the network, which is consistent with the gradual increase in training accuracy, validating the reasonableness of the assumptions made by Algorithm \ref{alg1}. DistrEE obtains an inference accuracy comparable to that of the non-early-exit method Last-exit, while the average computational overhead per image is saved by 27.6\%, and the inference latency is saved by about 78 ms. The experimental results show that the proposed DistrEE achieves an effective trade-off between inference latency and accuracy.

\begin{table}[!htbp]
\renewcommand\arraystretch{1.5}
  \begin{center}
    \caption{Performance for different early exit policies.}
    \begin{tabular}{c|c|c|c}
    \hline
    \multicolumn{4}{c}{DistrEE-2 (7 branches, 2 student models)}\\
    \hline
     Early exit policy & Accuracy(\%) & \begin{tabular}[c]{@{}c@{}}Computational \\ overheads (FLOPs)\end{tabular}  & \begin{tabular}[c]{@{}c@{}}Time delay\\ (ms)\end{tabular}\\
    \hline
    Last-exit & 93 & 34.26M & 1778.1 \\
    \hline
    DistrEE & 91 & 24.80M & 1700.2 \\
    \hline
    Random exit & 80 & 17.10M & 1493.3 \\
    \hline
    Similarity-based & 77 & 17.34M & 1569.9 \\ 
    \hline
    \end{tabular}
    \label{performance}
  \end{center}
\end{table}

\section{Conclusion}
In this work, we have presented an efficient distributed inference scheme called DistrEE, which adapts to the diversity of input data to improve inference performance by dynamically adjusting the inference plans of the devices. To achieve dynamic collaborative inference of deep learning models, we have integrated model early exit and distributed inference in DistrEE,  and designed offline joint training framework and online inference exit method for multi-branch models. Simulation results have shown that DistrEE can efficiently realize dynamic collaborative inference, maintain an inference accuracy comparable to non-early-exit methods, and significantly reduce inference computational overhead to achieve an effective trade-off between inference latency and accuracy.

\bibliographystyle{IEEEtran}
\bibliography{ref.bib}

\end{document}